\documentclass[aps,prd,twocolumn,superscriptaddress,groupedaddress]{revtex4}  

\usepackage{graphicx,units}
\usepackage{amssymb}
\usepackage{graphicx}
\usepackage{dcolumn}
\usepackage{float}
\usepackage{bm}
\usepackage{amsmath}

\makeatletter
\renewcommand*{\fnum@table}{{\normalfont\bfseries \tablename~\thetable}}
\renewcommand{\tablename}{Table}
\renewcommand*{\fnum@figure}{{\normalfont\bfseries \figurename~\thefigure}}
\renewcommand{\figurename}{Figure}
\makeatother

\usepackage{lipsum}

\usepackage{soul,xcolor}
\usepackage[normalem]{ulem}

\newcommand{\wb}[1]{\textcolor{yellow}}
\newcommand{\aht}[1]{\textcolor{black}{#1}}

\makeatletter
\renewcommand{\fnum@figure}{FIG. \thefigure}
\makeatother

\begin{document}

\preprint{APS/123-QED}

\title{Memory Effect or Cosmic String? Classifying Gravitational-Wave Bursts with Bayesian Inference}

\author{Atul K. Divakarla}
 \email{akd703@ufl.edu}
 \affiliation{%
 Department of Physics, University of Florida, 2001 Museum Road, Gainesville, FL 32611-8440, USA \\
 }
  \affiliation{%
 School of Physics and Astronomy, Monash University, Vic 3800, Australia \\
}%
\affiliation{
 OzGrav: The ARC Centre of Excellence for Gravitational Wave Discovery, Clayton VIC 3800, Australia
}%

\author{Eric Thrane}
 \email{eric.thrane@monash.edu}
 \affiliation{%
 School of Physics and Astronomy, Monash University, Vic 3800, Australia \\
}%
\affiliation{
 OzGrav: The ARC Centre of Excellence for Gravitational Wave Discovery, Clayton VIC 3800, Australia
}%

\author{Paul D. Lasky}%
 \email{paul.lasky@monash.edu}
 \affiliation{%
 School of Physics and Astronomy, Monash University, Vic 3800, Australia \\
}%
\affiliation{
 OzGrav: The ARC Centre of Excellence for Gravitational Wave Discovery, Clayton VIC 3800, Australia
}%

\author{Bernard F. Whiting}
 \email{bwhiting@ufl.edu}
 \affiliation{%
 Department of Physics, University of Florida, 2001 Museum Road, Gainesville, FL 32611-8440, USA \\
 }
 

\date{\today}

\begin{abstract}
\aht{
In the event of a gravitational-wave burst candidate, a key question will be: which astrophysical signal hypothesis is most likely?
Several different gravitational-wave transient sources can be modeled in the Fourier domain using a  simple power law.
}This power-law model provides a reasonable approximation for gravitational-wave bursts from cosmic string cusps, cosmic string kinks, and the memory effect.
Each of these sources is described using a different spectral index. 
In this work, we simulate interferometer strain data with injections of 
\aht{
memory and other power-law bursts
to demonstrate model selection in support of signal detection and for use in parameter estimation. 
}We show how Bayesian inference can be used to measure the power-law spectral index, thereby distinguishing between different astrophysical scenarios.
\aht{
We propose a strategy for model selection of power-law burst signals for gravitational-wave candidates, and we aim to use this analysis to determine whether a specific candidate can be best described by a compact binary coalescence (CBC) signal or by some other interesting astrophysical mechanism. 
}

\end{abstract}

\maketitle

\section{\label{sec:level1}Introduction}
The LIGO and Virgo Scientific Collaborations have cataloged eleven significant gravitational-wave signals in the first two observing runs ~\cite{LIGOScientific:2018mvr} 
\aht{
and released 56 possible candidates in the third observing run to date, 
}
each of them originating from the coalescence of binary black holes or neutron stars. 
Considerable effort is also undertaken to search for gravitational-wave ``bursts:'' unmodeled (or minimally-modeled) transients.
Bursting signals have been proposed for a variety of mechanisms including pulsar glitches ~\cite{PhysRevLett.87.241101}, neutron star collapse~\cite{Baiotti_2007}, core-collapse supernovae ~\cite{Abbott:2016tdt}, cosmic-string interactions ~\cite{PhysRevD.64.064008, PhysRevD.73.105001, Key:2008tt}, and gravitational-wave memory ~\cite{PhysRevLett.118.181103, PhysRevD.45.520}.
Previous searches for gravitational-wave bursts have included short-duration transients that arise from pulsar glitches and supernovae explosions ~\cite{PhysRevD.100.024017, PhysRevD.95.042003}, long-duration transients such as those from fallback accretion ~\cite{PhysRevD.99.104033, Abbott:2017muc}, cosmic strings ~\cite{Abbott:2017mem} , binary black hole mergers ~\cite{PhysRevLett.116.061102, PhysRevD.93.122004}, intermediate mass black hole mergers ~\cite{PhysRevD.96.022001}, post-merger remnants from neutron star mergers ~\cite{Abbott:2017dke}, sub-solar mass binaries  ~\cite{Authors:2019qbw}, eccentric binaries ~\cite{strathprints70364}, those associated with magnetar bursts ~\cite{Abbott:2019dxx}, those associated with neutrino emission ~\cite{PhysRevD.90.102002}, and those with electromagnetic counterparts ~\cite{PhysRevD.93.122008, Abbott:2016cjt}.
Active pipelines that search for unmodeled burst transients include coherent WaveBurst (cWB) ~\cite{Klimenko:2008fu}  and Omicron\textit{-}LALInferenceBurst (oLIB) ~\cite{PhysRevD.95.104046}.
The possibility of an unexpected source is the most promising motivation for gravitational wave burst searches.

\aht{
In the event of a gravitational-wave burst candidate, a key question will be: which astrophysical signal hypothesis is most likely?
To answer this question, gravitational-wave candidates are first identified by detection pipelines such as cWB ~\cite{Klimenko:2008fu}, PyCBC ~\cite{Usman:2015kfa}, and GstLAL ~\cite{Cannon:2011vi}. 
In the process of identifying, these detection pipelines utilize techniques such as time-slides to determine if the data contains an astrophysical signal or a detector glitch by time shifting the detector data around the trigger time; see ~\cite{Was:2009vh} for further details. 
Once the noise hypothesis is ruled out by these pipelines, implying that there exists some astrophysical signal in the data, parameter estimation analyses such as  Bilby ~\cite{Ashton:2018jfp} and LALInference ~\cite{Veitch:2014wba} begin by assuming Gaussian noise in the detectors and a CBC waveform approximant. 
In this paper, we are concerned with parameter estimation and model selection, and so we assume that a signal has already been detected by some other pipeline and that the noise is approximately Gaussian in the vicinity of the signal.
}

\aht{
By introducing the power-law model to the set of all astrophysical signals that are analyzed, our method can be used to classify between power-law signals in addition to CBC signals for parameter estimation of gravitational-wave data. 
Furthermore, we extend our analysis to show how we rule out power-law burst sources for gravitational-wave candidates that have their astrophysical source in question.
By combining our method with other burst classification techniques, it should be possible to construct a set of minimally-modeled bursts: supernovae bursts ~\cite{Logue:2012zw, PhysRevD.96.123013} , arbitrary power law bursts, an arbitrary superposition of wavelets ~\cite{Cornish:2014kda,Becsy:2016ofp}, etc.
Thus, our method contributes to the model selection for various astrophysical burst sources. 
By comparing the Bayesian evidence from the different models in the catalog, it will be possible to determine which catalog entry best describes the gravitational-wave burst candidate. 
}

This paper is structured as follows:
Section~\ref{sec:leve12} describes the set of power-law signal models used in the analysis. 
Section ~\ref{sec:level3} reviews the fundamentals of Bayesian inference, establishes the prior distribution for each signal parameter, and lists the steps taken to simulate and analyze interferometer strain data. 
Section ~\ref{sec:level4} presents results from a set of injection simulations with discussions on their implications. 

\section{\label{sec:leve12}Signal models}
We model signals with respect to phenomenological parameters, or signal characteristics, such that the dimensionality of each model remains low, ultimately reducing model complexity and improving statistical robustness. 
For example, rather than individually parameterizing the mass, distance, and inclination of an astrophysical system that emits memory, we just parameterize the ``amplitude'' and rise-time of a memory signal.
In all models, we use $t_A$ to denote the arrival time of a given burst signal. 

\subsection{\label{sec:level2A}Memory}
The non-linear memory effect, a prediction of general relativity, was first derived in ~\cite{PhysRevLett.67.1486}. 
The memory effect is a linearly polarized, DC, gravitational-wave signal originating from an anisotropic gravitational-wave energy flux; see ~\cite{Favata_2010}. 
Memory signals have a characteristic rise time $\tau$ proportional to the total mass of the astrophysical system ~\cite{PhysRevD.45.520}.
We model the time-domain waveform as, 
\begin{eqnarray}
h_{\text{m}} (t) = A_{\text{m}}  \tanh  \left(\frac{t - t_A}{\tau}\right) .
\end{eqnarray} 
Here $t$ is time, and $A_{\text{m}}$ is a parameter describing the memory amplitude.
In the Fourier domain, this memory model has the analytic form,
\begin{eqnarray}
\tilde{h}_{\text{m}} (f) =  - i \pi A_{\text{m}}  \tau e^{-2 \pi i f t_A} \left( \sinh (\pi^2 \tau f) \right) ^{-1} .
\end{eqnarray}
In the limit of $\tau \rightarrow 0$, corresponding to an astrophysical system with minimal mass, this memory model approaches $-i A_{\text{m}}  e^{-2 \pi i f t_A} /\pi f$, a power-law signal with a negative spectral index of one or equivalently, a step-function in the time domain. 

\subsection{\label{sec:leve2B}Cosmic String Interactions}
It has been theorized that topological defects during a symmetry-breaking phase transition in the early Universe can give rise to one-dimensional strings that expand to cosmological scales and form cosmic string networks ~\cite{PhysRevD.64.064008}.
A network of cosmic strings can produce 
\aht{
linearly polarized
}gravitational-wave bursts, arising from features such as cusps ~\cite{Key:2008tt} and kinks through string interactions.
Cusps are parts of a cosmic string that move at relativistic speeds, their gravitational-wave emission is modelled by,
\begin{eqnarray}
\tilde{h}_{\text{c}} (f) = A_{\text{c}} e^{-2 \pi i f t_A} f^{-4/3} \Theta \left(f_h - f \right).
\label{cuspeq}
\end{eqnarray}
\aht{
Here and below, $f_h$ is a high frequency cutoff parameter that is inversely proportional to the cube of the beaming angle, which is defined as the angle between the line of sight and emission cone axis; see ~\cite{Key:2008tt} for further details.
}Kink signals are sourced from discontinuities in the string's tangent vector, with their resulting waveform being modelled as,
\begin{eqnarray}
\tilde{h}_{\text{k}} (f) = A_{\text{k}} e^{- 2 \pi i f t_A} f^{-5/3} \Theta \left(f_h - f \right).
\label{kinkeq}
\end{eqnarray}

\begin{figure*}
  \centering
   \includegraphics[width=\textwidth,height=9cm]{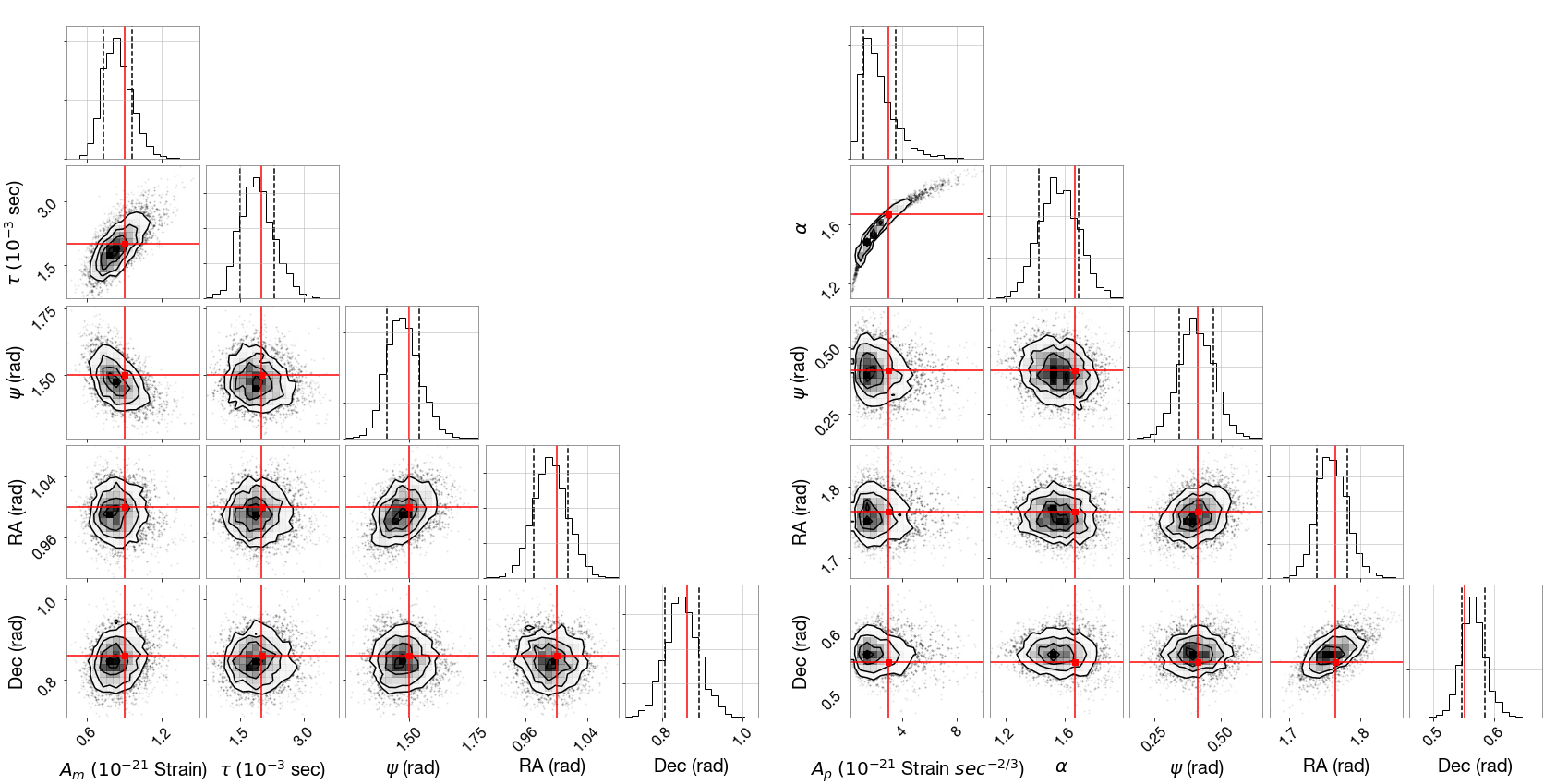}
  \caption{
  Marginalized posterior distributions for each signal parameter of the memory model (left), with signal/noise log Bayes factor $\ln(\text{BF})$ = 68.8 and 
  \aht{an arbitrary power-law model (right), with $\alpha = $ 5/3, and with signal/noise $\ln(\text{BF})$ = 60.8. }The red lines show the values of the injected parameters and the dotted lines in each one-dimensional posterior distribution represent one standard deviation credible intervals. The contour levels of the two-dimensional posterior distributions are $0.5, 1.0, 1.5, \text{and} ~ 2$ standard deviations. These plots show how well we recover signal parameters of a given astrophysical model.
  } 
  \label{fig:corners}
\end{figure*}

\subsection{Arbitrary Power Law}
Inspired by the morphological similarity between memory and string waveforms we propose to search for a general power-law signal, which might capture a range of scale-invariant phenomena that emit gravitational waves.
We assume linear polarization and define a general power-law model, 
\begin{eqnarray}
\tilde{h}_{\text{p}} (f) = A_{\text{p}} e^{- 2 \pi i f t_A} f^{-\alpha},
\label{pleq}
\end{eqnarray}
where $\alpha$ is the power-law spectral index.
\section{\label{sec:level3}Method}

\subsection{Bayesian Inference}
We use Bayesian inference to estimate parameters and calculate the significance of each signal model given observed data.  
We denote the complex, frequency-domain strain data to be $d$ and a given frequency-domain waveform model $\mu(\theta)$ as a function of signal parameters $\theta$.
We denote the prior distribution on signal parameters as $\pi(\theta)$.
We then construct the posterior probability distribution $p (\theta | d)$ using Bayes' theorem
\begin{eqnarray}
p \left(\theta | d \right) = \frac{\mathcal{L} \left(d | \theta \right) \pi (\theta)}{\mathcal{Z}}.
\end{eqnarray}
We employ a Gaussian likelihood function,
\begin{eqnarray}
\ln \mathcal{L}  \left(d | \theta \right) = - \frac{1}{2} \Big[ \sum_j \frac{|d_j - \mu_j (\theta)|^2}{\sigma_j ^2} + 2 \ln \left(2 \pi \sigma_j ^2 \right) \Big].
\end{eqnarray}
The summation in the likelihood function is taken over $j$ frequency bins and $\sigma_j$ is the noise amplitude spectral density of the detector in the $j^{th}$ frequency bin. 
We assume Gaussian noise for the remainder of our analysis. 
The evidence is calculated by
\begin{eqnarray}
\mathcal{Z} =  \int \mathcal{L} \left(d | \theta \right) \pi (\theta) d \theta.
\end{eqnarray}

\aht{
Given a pair of signal models, $M_{1}$ and $M_{2}$ with the ratio of the prior odds set to unity, we compare models to noise $N$, and models to each other by computing Bayes factors
$\text{BF}^{M_{1}} _ {N} = \mathcal{Z}_{M_1} / {\mathcal{Z}_N}$ (signal versus noise) and $\text{BF}^{M_1} _ {M_2} = \mathcal{Z}_{M_1} / \mathcal{Z}_{M_2} $ (model one versus model two).
}Following convention~\cite{Jeffreys61}, we here consider $\ln(\text{BF})>8$ to be strong evidence for a particular hypothesis.

\subsection{Prior Distributions}
\aht{
We establish the choice of prior distribution for each parameter present in each of the signal models.
The bounds of our prior distributions are chosen to match the observing band of the detectors. For example, $\pi (\tau)$ and $\pi (f_h)$ are uniformly distributed with bounds from $\unit[0.5-50]{ms}$ and $\unit[20-2000]{Hz}$, respectively. 
We chose a log uniform prior for the amplitude of the first cosmic string, memory, and power law detection to reflect our belief that any order or magnitude is as likely as any other order of magnitude.
We chose the prior for the spectral index in Eq. \ref{pleq}, $\pi (\alpha)$, to be uniform on the interval $[0,2]$,
which includes the cosmic string models without the frequency cutoff as a subset of the power-law model.
}We use standard priors for extrinsic parameters such as polarization angle and sky location.
We explicitly marginalize over the arrival time and we can numerically reconstruct its posterior distribution in post-processing; see, e.g.,~\cite{Thrane:2018qnx}.

\begin{figure*}
  \centering
  \includegraphics[width=\textwidth,height=7cm]{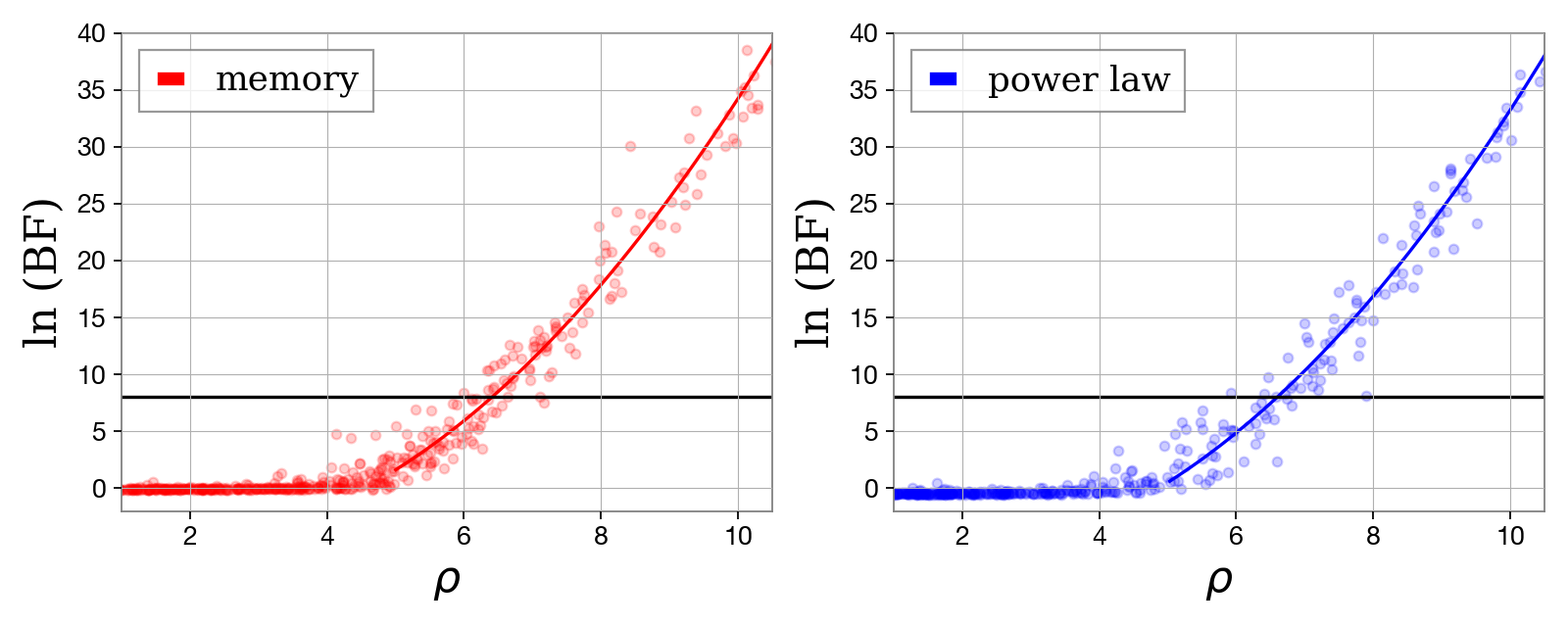}
  \caption{
  Signal/noise Bayes factor versus matched-filter signal-to-noise ratio $\rho$
  for simulated memory (left), and power-law events (right). 
  The signal model is chosen to match the injection.
  The horizontal axis is the matched filter signal-to-noise ratio.
  The individual dots represent the Bayes factors for a particular injection.
  We fit a quadratic function, shown by the solid curves, to each panel. 
  The horizontal black line corresponds to a Bayes factor of $\ln \text{BF} = 8$.
  These plots quantify how loud a memory or power-law burst signal has to be for the signal hypothesis to be significantly larger than the noise hypothesis. 
  }
  \label{fig:bfvssnr_all}
\end{figure*}

\subsection{\label{sec:proced}Procedure}

We outline the steps taken to simulate interferometer strain data and to run our Bayesian inference analysis on a given signal injection. 
We set the total duration of the data segment to be 4 seconds at a sampling rate of 4096 Hz. 
All signals are injected with $t_A = \unit[2]{s}.$
We assume a three-detector network consisting of two Advanced LIGO detectors and one Advanced Virgo detector operating at design sensitivity~\cite{Aasi:2013wya}.
We generate Gaussian noise and inject a power-law signal with a randomly chosen set of injection parameters. 
We set the minimum and maximum frequency of our detectors to be $[\unit[20]{Hz},\unit[2000]{Hz}]$, corresponding to the most sensitive region of our detectors.
We define the total joint matched-filter signal-to-noise ratio (SNR) and inner-product to be
\begin{align}
\rho & = \frac{\langle d, \mu \rangle_\text{tot}} {\langle\mu,\mu\rangle_\text{tot}^{1/2}} , \\
\big \langle a,b \big \rangle_\text{tot} & \equiv 4 \Delta f \sum_k \sum_{j} \mathcal{R} \left( \frac{a_{j,k} ^{*} b_{j,k} }{\sigma^2 _{j,k} } \right) .
\end{align}
Here, $\Delta f$ is the frequency resolution 
and $k$ is an index for the detector. 
We carry out inference using the nested-sampling algorithm ``dynesty'' with 500 live points~\cite{2019arXiv190402180S}.
We use the Bayesian inference software package Bilby ~\cite{Ashton:2018jfp} to perform our analysis.

\section{\label{sec:level4}Simulations}

\subsection{Parameter Estimation}
To infer the properties of a gravitational-wave burst event, we calculate the posterior distributions for each parameter in a given signal model. 
We show the one-dimensional and two-dimensional posterior distributions for an injection of a memory and cosmic string kink signal in Fig.~\ref{fig:corners}, quantifying how well we recover a set of injected parameters.
\aht{
The posterior distribution for the spectral index in the power-law model differentiates between various astrophysical burst scenarios.
}

\subsection{Signal Detection}
\aht{
We calculate how loud a memory and power-law event must be for the respective signal hypothesis to be significantly more likely than the noise hypothesis. 
}We show results from injection simulations of memory and power-law signals in Fig.~\ref{fig:bfvssnr_all}.

\subsection{Model Selection}
\begin{figure}[H]
    \centering
    \includegraphics[height = 6.0cm, width=8.6cm]{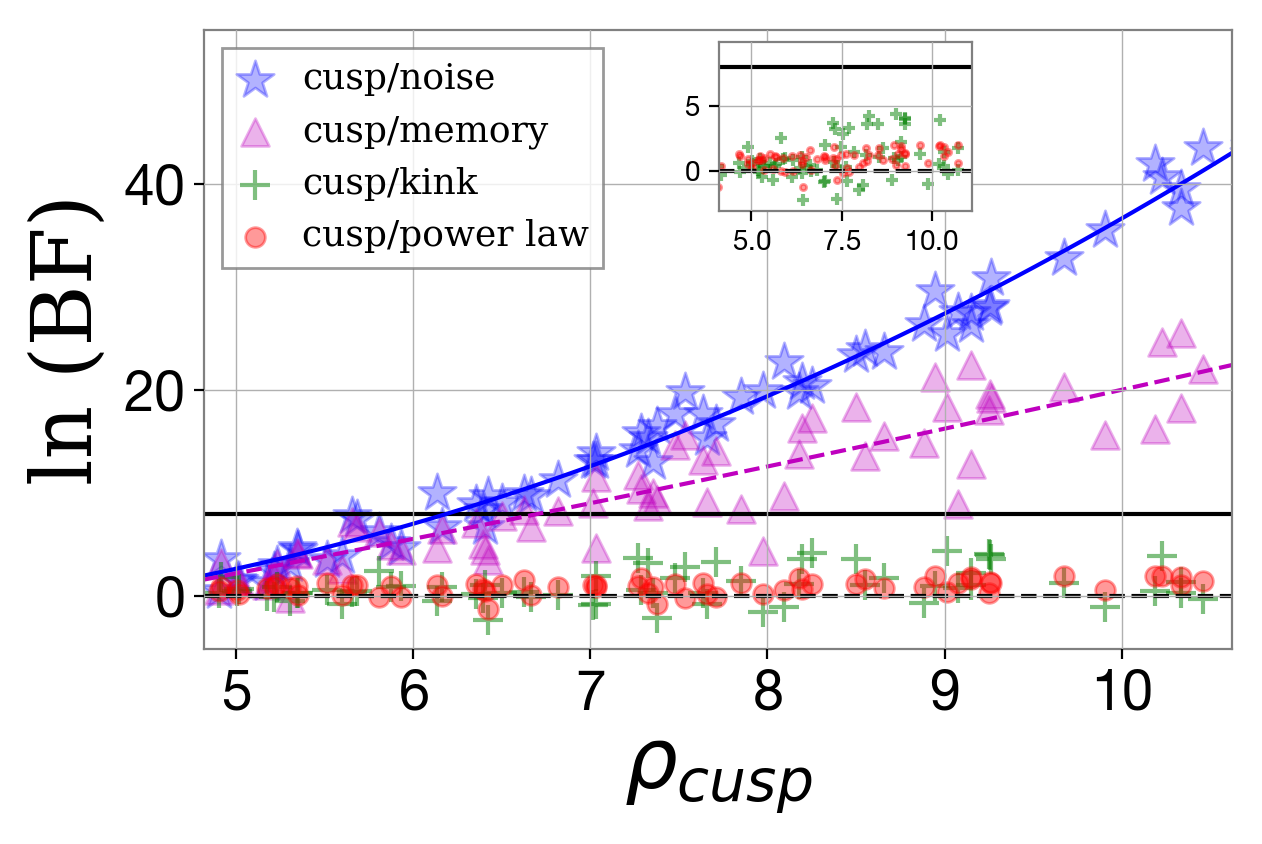}
    \caption{
    In these simulations, the individual points represent the calculated Bayes factors when comparing the cusp signal model to noise (blue stars), memory (magenta triangles), kink (green plus) and power-law (red circles) models.
    Quadratic fits comparing cusp to noise (solid blue) and cusp to memory (dashed magenta) are shown.
    The horizontal black solid line corresponds to $\ln(\text{BF}) = 8$ and the dashed line to $\ln(\text{BF}) = 0$. This plot establishes how well we can distinguish cosmic string bursts from other power-law bursts. 
    }
    \label{fig:pl_mem}
\end{figure}

Given that the noise hypothesis is ruled out by one of the detection pipelines, we can also differentiate between various astrophysical burst scenarios by calculating the model versus model Bayes factor for a pair of signal hypotheses.
\aht{
Suppose we were to detect a cosmic string cusp signal, how do we establish that it is not a memory or kink burst? 
To demonstrate how we use the Bayes factor in such a scenario, we simulate a set of cosmic string cusp signal events with $f_h = \unit[1000]{Hz}$ and perform a search for a cusp, kink, memory and power-law burst for each individual simulation. 
By incrementally increasing the amplitude of the cusp signal throughout our injections, while keeping all other parameters fixed , we observe how our confidence increases with the loudness of the injected cusp signal, shown in Fig.~\ref{fig:pl_mem}. 
}

\aht{
When comparing the cusp to memory at a signal-to-noise ratio of $\sim 7$, the cusp hypothesis is more significant than the memory hypothesis. 
The cusp/power law Bayes factor remains between $\ln(\text{BF}) = 0$ and $\ln(\text{BF}) = 8$ in Fig.~\ref{fig:pl_mem}, which implies it is difficult to significantly distinguish between a cosmic string signal with a large frequency cutoff and an arbitrary power-law signal in this signal-to-noise ratio interval. 
}

\subsection{Extension to Binary Black Hole Mergers}

\begin{figure}[H]
    \centering
    \includegraphics[height = 6.5cm, width=8.6cm]{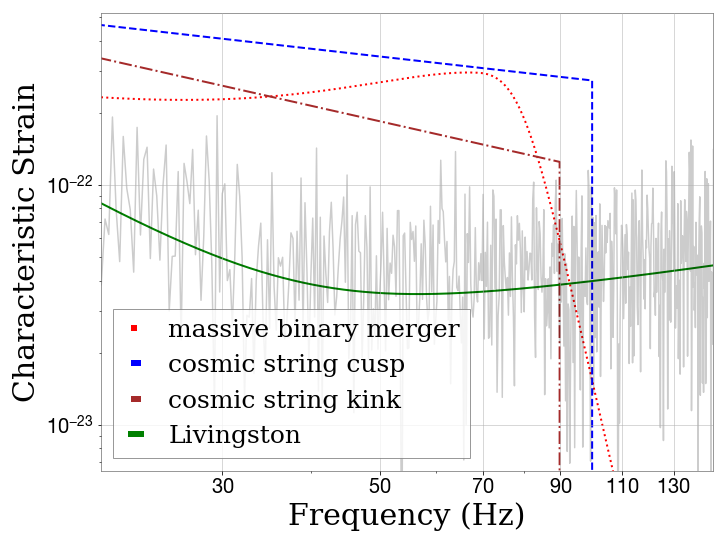}
    \caption{
     Frequency domain plot showing the characteristic strain of the noise amplitude spectral density of the Livingston detector (solid green curve) and a noise realization (gray region), a massive binary merger with $M_{\text{tot}} = \unit[250]{ M_{\odot} }$ (dashed blue curve), a cosmic string cusp (dotted red curve), and a cosmic string kink (dotted dashed brown curve). 
     When injected into a three-detector network, all signals have similar signal-to-noise ratios. }
    \label{fig:bbh_cusp}
\end{figure}

\aht{
We now expand our set of bursting signals to include a CBC signal and calculate the model-to-model Bayes factors for a CBC signal and some other power-law burst.
For example, in the event of a detection of a massive binary black hole merger, one with large component spins, or one with a large mass ratio, the merger waveform may look qualitatively different than the ``chirping'' waveforms observed thus far ~\cite{LIGOScientific:2018mvr}. 
In the case of a massive binary black hole merger, the CBC signal can look similar to a cosmic string cusp or kink signal, shown in Fig.~\ref{fig:bbh_cusp}.
}In such situations, it will be natural to ask whether or not a particular gravitational-wave candidate originated from a compact binary coalescence or a cosmic string interaction.  

As an example of model selection between CBC and cosmic string signals, we simulate an equal-mass massive binary black hole merger signal with total mass $M_{\text{tot}} = \unit[250]{ M_{\odot} }$, at signal-to-noise $\rho = 14.5$, using the IMRPhenomPv2 waveforms  ~\cite{PhysRevD.93.044006,
PhysRevD.93.044007}. 
This results in a signal that is in-band for only the merger-ringdown portion of the coalescence. 
We perform a search for a binary black hole, a cosmic string cusp (Eq. \ref{cuspeq}), and a cosmic string kink (Eq. \ref{kinkeq}) signal. 
We calculate the following Bayes factors: $\ln (\text{BF}^{BBH}_{cusp}) = 86.9 ~\text{and} ~ \ln (\text{BF}^{BBH}_{kink}) = 86.7 $. Given the magnitude of these Bayes factors, the black hole signal hypothesis is significantly more likely than either of the cosmic string burst signals.
In this case, we are able to rule out cosmic strings as the astrophysical origin of the burst event.


\section{\label{sec:level5}Summary and Outlook}

In this paper, we present a unified framework for classifying gravitational-wave burst signals from memory, cosmic strings, and other bursts that are well approximated by a power law in the Fourier domain. 
\aht{
We establish how the arbitrary power-law models can be used to search for signals that cover a wide range of astrophysical burst scenarios.
}Using simulations, we estimate the signal parameters of a gravitational-wave burst, identify astrophysical signals from interferometer noise, and characterize between power-law bursts by measuring their spectral index.

In testing our classification method we have assumed Gaussian noise.
In reality, ground-based interferometer detector noise is populated by glitches, power lines, seismic activity, and other transient noise sources.
In future work, we will investigate how to handle the non-Gaussian nature of noise ~ \cite{Ashton:2019wvo}, for example, using techniques such as the Bayes Coherence Ratio defined in ~\cite{Isi:2018vst}. 

Our signal models are of low dimensionality, especially when compared to the large parameter space of CBC signals. 
Therefore, it takes only a few minutes to analyze a few seconds of data using a single core processor.
\aht{
We aim to use this analysis to determine whether a gravitational-wave candidate can be best described by a CBC signal or by some other interesting astrophysical mechanism. 
}


\begin{center}
\textbf{Acknowledgements}
\end{center}
We thank Colm Talbot, Rory Smith, and Gregory Ashton
who provided helpful discussion of parameter estimation and Bayesian statistics.
The authors gratefully acknowledge the support from the University of Florida through the National Science Foundation (NSF) grants PHY-1205512, PHY-1460803, PHY-1607323 and the Australian Research Council (ARC) grants CE170100004, FT150100281, FT160100112, and DP180103155.
\aht{Computing resources of the LIGO Data Grid at the Caltech Institute of Technology were used for this paper.}
We thank the 2018 Australia-Americas PhD Research Internship Program, which was supported by the Australian Government Department of Education and Training through the Australian International Education: Enabling Growth and Innovation (EGI) project fund, for further information, please visit www.internationaleducation.gov.au.’
This work was also supported by Monash University in providing resources and accommodation. 
This is LIGO document No: LIGO-P1900297.

\bibliographystyle{plain}
\bibliography{main.bib}
\bibliographystyle{unsrt}

\end{document}